\documentclass[preprint]{aastex}

\slugcomment{PASP, in press}

\begin{document}

\title{A Computational Tool to Interpret the Bulk Composition of Solid Exoplanets based on Mass and Radius Measurements}

\author{Li Zeng\footnote{Department of Physics, Massachusetts Institute of Technology, 77 Massachusetts Ave., Cambridge, MA  02139 },  S. Seager$^{1,}$\footnote{Department of Earth, Atmospheric
and Planetary Sciences, Massachusetts Institute of Technology, 77 Massachusetts Ave., Cambridge, MA  02139 }}

\begin{abstract}
The prospects for finding transiting exoplanets in the range of a few
to 20 $M_{\oplus}$ is growing rapidly with both ground-based and
spaced-based efforts.  We describe a publically availalble computer
code to compute and quantify the compositional ambiguities for 
differentiated solid exoplanets with a measured mass and radius,
including the mass and radius
uncertainties.
\end{abstract}

\section{Introduction}

Over 250 extrasolar planets are known to orbit nearby main sequence
stars.  Among
these include over a dozen exoplanets with minimum masses below 22
$M_{\oplus}$ and several with minimum masses less than 10 $M_{\oplus}$. Of
key interest are transiting planets with measured masses and radii,
which can be used to constrain the planet's interior bulk composition.
The relationship between mass and radius for solid exoplanets has
hence received much attention in the last few years \citep{vale2006,
fort2007, seag2007, sels2007, soti2007}. The recent
activity builds on much earlier work
\citep{zs1969, stev1982}, with improvements on the equations of state
and treatment of different mantle and core compositions to varying
degrees of complexity.

Unlike for the solar system planets, we have no access to the
gravitational moments of exoplanets. Hence the density distribution in
the interior is unknown and this leads to an ambiguity, or degeneracy,
in the interior composition for an exoplanet of a fixed mass and
radius.  One way to capture the degeneracies of exoplanet interior
composition is using ternary diagrams (introduced to exoplanet
interiors by \citet{vale2007}).

We adopt the idea of using ternary diagrams to quantify the
compositional uncertainty in exoplanets.  The planet mass and radius
are the observed quantities and therefore we focus solely on on
ternary diagrams for a planet of fixed mass and fixed radius
\citep[c.f.][]{vale2007}.  We compute ternary diagrams for solid
exoplanets ranging in mass from 0.5 to 20 $M_{\oplus}$. We
furthermore explain the behavior of the mass-radius curves in two and three
dimensions. We also present a description of our publically available
computer code to compute fixed mass-radius ternary diagrams, including
the observational uncertainties.

\section{Computer Model}

\subsection{Background and Equations}
\label{sec-eqns}
We begin by assuming the major components of a solid exoplanet are
limited to an iron core,  a silicate mantle, and a water ice outer layer. In other words, we
assume the interior of the planet is differentiated with the
denser materials interior to the less dense materials. We further assume
each layer to be homogeneous in its composition. 

We can then model the interior of a solid exoplanet by using:
 
(1)	the equation for mass of a spherical shell 
\begin{equation}
\frac{dm(r)}{dr}
=4 \pi r^2 \rho(r),
\end{equation}
where $m(r)$ is the mass included in radius $r$, $\rho(r)$ is the density at radius $r$;

(2) the equation for hydrostatic equilibrium
\begin{equation}
\frac{dP(r)}{dr}=
\frac{-Gm(r) \rho(r)}{r^2},
\end{equation}
 where $P(r)$ is the pressure at radius $r$;
and

(3) the equation of state (EOS) that relates $P$ and $\rho$.

The EOS is different for each different material. We used Fe ($\epsilon$)
for the planet core, MgSiO$_3$ perovskite for the silicate mantle,
and water ice VII, VIII, and X for the water-ice outer layer. 
See \citet{seag2007} for a detailed discussion of the EOSs
including their source.  The temperature has 
little effect on the EOS especially in the high pressure regime \citep{seag2007}; 
we ignore the temperature
dependence of the EOS. This simplifies the equations and their solution,
while enabling a relatively accurate analysis.

In this problem, we have five variables: 
\begin{enumerate}
\item	the iron mass fraction ($\alpha$);
\item	the silicate mass fraction ($\beta$);
\item	the water-ice mass fraction ($\gamma$);
\item	the total mass of the planet ($M_p$);
\item	and the mean radius of the planet ($R_p$). 
\end{enumerate}
The variables $\alpha$, $\beta$, and $\gamma$  are not 
independent of each other. Based on the assumption
that the planet only consists of iron, silicate, and water we have:
$\alpha + \beta + \gamma = 1$, which can also be expressed as $\gamma
= 1 - \alpha - \beta$. We therefore have four variables:
$\alpha, \beta, M_p$ and $R_p$. Given any three of
these variables, we can determine the fourth variable uniquely.
We can also see that given an $M_p$ and $R_p$, there is a
relationship between $\alpha$ and $\beta$. There is not a single value
of $\alpha$ and $\beta$ that produces a given $M_p$ and
$R_p$. Instead, there are infinite pairs of $\alpha$ and $\beta$
which give the same $M_p$ and $R_p$, and we call this a
degeneracy in the interior composition.

\subsection{Algorithm for Solving the Differential Equations}
\label{sec-alg1}
Our program integrates from the surface $r = R_p$ inward to the center
of the planet. The outer boundary condition is $m(R_p)=M_p$ and
$P(R_p)=0$. That is, at the surface,  the mass is the specified total
planet mass and the pressure is approximately 0.

We aim to interpret observations of a planet of a given mass and
radius.  We therefore choose to intergrate inwards instead of outwards
based upon the known parameters of the planet ($M_p$, $R_p$, and
$P(R_p)$).  The independent variable is $r$, decreasing from $r=R_p$
to $r=0$. The interior boundary condition is $m(r) = 0$ at $r=0$.
Typically, when integrating $m(r)$, $m$ does not equal zero at
$r=0$. We therefore must iterate, tuning the mass fraction of each
layer until $m=0$ and $r=0$ is reached.

Given a single $M_p$ and $R_p$ the computer program finds all possible
combinations of ($\alpha$, $\beta$, $\gamma(=1-\alpha-\beta)$) that
give the same planet mass and radius. The program begins with a chosen
value of $\alpha$, and then takes a guess of $\beta$. Next, the
computer program integrates the differential equations (1) and (2)
using the above boundary conditions to find the value of the planet
radius. By comparing this radius to the desired $R_p$, the computer
program tunes the value for $\beta$, by using the bisection
method. This process is repeated several times, until $\beta$ is found
to a satisfactory accuracy of 1/1000.  By varying $\alpha$ within the
range of 0 to 1, we can get all possible combinations of $\alpha$ and
$\beta$ which produce a specific $M_p$ and $R_p$.

\subsection{Algorithm for Generating a Database of $M_p = 0.5$Ð-$20 M_{\oplus}$ }
\label{sec-alg2}
$M_p$ and $R_p$ are observed parameters and one usually wants to find
the corresponding allowed $\alpha$ and $\beta$. For a range of $M_p$
and $R_p$---corresponding to observational uncertainties---it can be
very time consuming to use the first algorithm described in
\S\ref{sec-alg1}. We therefore generate a database that is a discrete
representation of the relation $R_p = R_p(\alpha, \beta,
M_p)$. Figure~\ref{fig:3D} illustrates the 4D database.

This database is a 3-D array which contains the data of $R_p$
corresponding to each combination of $\alpha$ and $\beta$ (0 to 1 with
1\% spacing) and $M_{\oplus}$ (ranging from 0.5 to 20 $M_{\oplus}$
with $0.25 M_{\oplus}$ spacing). The database can be used via linear
interpolation to find $R_p$ for any given $M_p$, $\alpha$ and
$\beta$. A conservative estimate of the fractional error in the
database interpolation is 1/1000.  For the same range of $M_p$ and
$R_p$, interpolation in the database is about 45 times faster than
solving the differential equations.

To generate a database of all values of $\alpha$, $\beta$, and $R_p$
for $M_p$ ranging from 0.5 to 20 $M_{\oplus}$.  Given $M_p$, $\alpha$,
and $\beta$, this algorithm also integrates from the surface inward to
$r=0$ and $m(r)=0$ to find $R_p$. In contrast to the first algorithm
(which solves for a given $M_p$ and $R_p$), a single integration in
radius results in the desired solution of $R_p$, for a given $M_p$,
$\alpha$, and $\beta$. In other words is there is no iteration
required, making this algorithm much more efficient.

\subsection{Instructions for Downloading and Using the Code}

The code is based in MATLAB and can be downloaded from
http://web.mit.edu/zengli/www/ under ``Research Field'' or from
http://seagerexoplanets.mit.edu/research/interiors.html If using this
computer code please cite this paper and also \citet{seag2007}.

We have made two different codes available. The codes have the same
output, but the first is based on a differential equation solver
(\S\ref{sec-alg1}) and the second code is based on interpolation of
the large database (\S\ref{sec-alg2}).  For the codes, the planet mass
must be in the range 0.5--20~$M_{\oplus}$. The inputs to the codes
are: the planet mass in Earth masses ($M_p$), the planet mass
uncertainty in Earth masses ($\sigma_{M_p}$), the planet radius in
Earth radii ($R_p$), and the planet radius uncertainty in Earth radii
($\sigma_{R_p}$). The values $\sigma_{M_p} = 0$ and $\sigma_{R_p}$ = 0
are allowed. If the combination of input values $M_p$ and $R_p$ are
unphysical, the code will return an error.

{\bf ExoterDE($M_p$, $\sigma_{M_p}$, $R_p$, $\sigma_{R_p}$).}  This
code solves the two differential equations described in
\S\ref{sec-eqns}. This code consists of three subroutines (each of
which must be downloaded) that are automatically called by the above
command. The first subroutine is the differential equation solver,
which also reads the equations of state.  The second subroutine
contains the actual differential equations. The third subroutine plots
the ternary diagrams; this subroutine calls a ternary diagram plotting
routine \citep{the2005}\footnote{
http://www.mathworks.com/matlabcentral/\\
fileexchange/loadFile.do?objectId=7210\&objectType=file} which plots a
single line for each of the 1-, 2-, and 3-$\sigma$ contour lines. See
\S \ref{sec-uncertainty}.  An example from this code is shown in
Figure~\ref{fig:ternunc}.

{\bf ExoterDB($M_p$, $\sigma_{M_p}$, $R_p$, $\sigma_{R_p}$)}.  This
code reads in the database of $M_p$, $R_p$ and fractional composition
($\alpha$ and $\beta$). The output is a ternary diagram, shaded
throughout the 1-, 2-, and 3-$\sigma$ contour curves.  This subroutine
uses the same ternary diagram plotting routine as described above.

The differential equation solver ExoterDE is much slower than the
database extracter ExoterDB. In principle, ExoterDB is more accurate
than ExoterDB.

\section{Data Display}

\subsection{2-D Cartesian Diagram}
For a given $M_p$ and $R_p$ we want to know the interior composition
of the relative mass fraction of the three components. There are three
variables we have solved for ($\alpha, \beta, \gamma$), but only two
of them are independent (since $\gamma=1-\alpha-\beta$). Therefore
points on a 2-D diagram can describe all the possible combinations of
$\alpha, \beta, \gamma$ for a given $M_p$ and $R_p$. We show such a
solution in Figure~\ref{fig:cart1}. We note that $0 \leq \alpha
\leq1$, $0 \leq \beta \leq 1$, $\alpha + \beta \leq 1$, and therefore
not every point in the 2-D plane will correspond to a set of $\alpha,
\beta, \gamma$. Only the points which are in a right-angled triangular
region will respresent the set of allowed solutions.

\subsection{Ternary Diagram}

Ternary diagrams to describe the interior composition of exoplanets
were introduced by \citet{vale2007}. In a ternary diagram, $\alpha$,
$\beta$, and $\gamma$ are each one axis of an equilateral
triangle. Although $\gamma$ is extraneous, the ternary diagram is
useful because it is more intuitive to see the three components of the
planet interior (in a symmetric way) compared to a 2D Cartesian
diagram with only two of the components. Figure~\ref{fig:terndef}
shows how to read a ternary diagram.
 
\subsection{Relationship Between the 3D  and 2D Cartesian Diagrams and the Ternary Diagram}
\label{sec-convert}
To explain the full origin of a curve on the ternary diagram we start
with the 3D Cartesian diagram with all solutions of $M_p$, $R_p$ and
composition (in terms of $\alpha$ and $\beta$), as shown in
Figure~\ref{fig:3D}.  We take an isoradius and isomass surface as
shown in Figure~\ref{fig:isorisom}.  As an example, in
Figure~\ref{fig:isorisom} the red surface is the iso-radius surface of
$R_p = 1.7 R_{\oplus}$ and one of the blue colored planess is the
iso-mass surface of $M_p = 4 M_{\oplus}$. These two surfaces intersect
each other and result in a curve. This curve can be projected
vertically to the $x$-$y$ plane which is the iso-mass plane. We
therefore have a (iso-radius and isomass) curve on the isomass
plane. This curve is shown in Figure~\ref{fig:cartproj} in a Cartesian
diagram.

The Cartesian and ternary diagrams are two different ways to represent
the same information. There exists a linear coordinate transformation
between the two. That means if a function is a straight line appearing
in the 2-D Cartesian diagram, it will still be a straight line in the
ternary diagram.

The transformation from 2-D Cartesian coordinates to the ternary
diagram coordinates is
\begin{equation}
\label{eq:coordconversiona}
x_{ternary}= \frac{1}{2} (1+x-y)
\end{equation}
\begin{equation}
y_{ternary}=\frac{\sqrt{3}}{2} (1-x-y).
\label{eq:coordconversionb}
\end{equation}
Here $x$ and $y$ are the coordinates of a point in a 2D Cartesian
diagram, the $x_{ternary}$ and $y_{ternary}$ are the coordinates of the point
in ternary diagram  in the Cartesian grid variables.
Figure~\ref{fig:ternproj} shows the same $R_p = 1.7 R_{\oplus}$, $M_p = 4
M_{\oplus}$ curve represented by a ternary diagram.

\section{Results and Discussion}
\subsection{Observational Uncertainties}
 \label{sec-uncertainty}
Real planet mass and radius measurements have uncertainties.  The
planet mass and radius uncertainties are typically 5 to 10 percent
\citep[e.g.,][]{sels2007}, and even smaller for the most favorable targets.  We
now present examples of ternary diagrams that include the mass and
radius uncertainties.

We consider uncertainties of 1, 2, and 3 standard deviations
($\sigma$) from the measured value.  See Figures~\ref{fig:ternunc} and
\ref{fig:ternmult}. In more detail, the uncertainty in composition on
the ternary diagram is
\begin{equation}
\sigma_{comp}=\sqrt{(\sigma_{compM})^2+(\sigma_{compR})^2},
\end{equation}
where $comp$ refers to composition and $compM$ and $compR$ refer to
composition uncertainties caused by the planet mass and radius
uncertainty respectively. Here we have assumed that the uncertainties
in mass and radius are independent from each other and have assumed
the linearity of the superposition of small uncertainties. 

Figure~\ref{fig:contour} shows a planet with $M_p = 10 \pm 0.5
M_{\oplus} $ and $R_p = 2 \pm 0.1 M_{\oplus}$. We see that taking the
3-$\sigma$ limit, almost the entire ternary diagram is filled. In
other words, for a 5\% 3-$\sigma$ (i.e., 15\%) uncertainty on the
planet mass and radius, the interior composition in terms of
fractional composition of iron, silicates and water cannot be
determined.  The reason this example fills the almost the whole
ternary diagram is that a planet with $10 M_{\oplus}$ and $2
R_{\oplus}$ has an average density in between two extreme cases
(purely iron or purely water). Therefore, a large variety of different
combinations of iron, silicate and water can result in a similar $M_p$
and $R_p$.  Even taking a 1-$\sigma$ uncertainty of the planet mass
and radius, the uncertainty in internal composition is large.

We note that an uncertainty in $R_p$ has more of an effect on the uncertainty in the interior composition than an uncertainty in $M_p$. This is because the planet's average density
$\rho \sim M_p/R^3$. Considering error propagation, the uncertainty in radius 
has a three times larger effect on the uncertainty in average density than does
the mass uncertainty.

We show ternary diagrams for planets with various masses, radii, and
5\% fractional uncertainty in Figure~\ref{fig:ternmult}.  Only
solutions in part of the ternary diagram are allowed, despite
considering the 3-$\sigma$ range. In Figure~\ref{fig:ternmult}b, the
upper 3-$\sigma$ boundary is absent because it goes below the lowest
allowed density of the $2M_{\oplus}$ and $1.5 R_{\oplus}$ planet and
is thus unphysical. In Figure~\ref{fig:ternmult}c we see the opposite
case, where the lower 3-$\sigma$ boundary is absent because it goes
above the highest planet density allowed and is thus unphysical.

A ternary diagram for a fixed planet mass and radius that includes
observational uncertainties is one of the primary outcomes of this
paper.

\subsection{Model Uncertainties}

The model and computer code we present assumes a differentiated planet
composed of an iron core, a silicate mantle, and a water ice outer layer.
The division into three major materials is 
based on the point that the densities of iron, silicate,
and water are much more different from each other than any
minor compositional variant of each individual material. 
The model neglects phase variation and temperatures which, as argued
in \citet{seag2007}, have little effect on the total
planet radius (to an uncertainty of about 
$\sim$1 to 3 percent uncertainty in planet radius, decreasing with 
increasing planet mass.). 

Low-pressure phase changes (at $< 10$ GPa) are not important for a
planet's radius because for plausible planet compositions most of the
mass is at high pressure.  For high pressure phase changes we expect
the associated correction to the equation of state (and hence derived
planet radii) to be small because at high pressure the importance of
chemical bonding patterns to the equation of state drops.

Regarding temperature, at low pressures ($\lesssim$ 10 GPa) in the outer planetary layers,
the crystal lattice structure dominates the material's density and the
thermal vibration contribution to the density are small in comparison.
At high pressures the thermal pressure contribution to the EOS is
small because the close-packed nature of the materials prevents structural
changes from thermal pressure contributions.  

 Although the code can model radii for planets in the mass range
 0.5  to 20~$M_{\oplus}$, the model is more accurate for planets
 above  a few Earth masses \cite{seag2007}.

 The model also neglects variation in composition, such as a light element
 in the iron core as Earth and Mercury are believed to have. The model also
 omits other impurities in the mantle and water layer, including
 iron in the mantle. Molten cores have
 also been omitted. At the present time, these model uncertainties are expected
 to have an effect on the planet radius much less than the $\sim$5 percent
  radius observational uncertainty.
  
 For all of the above reasons, we therefore argue that for the present
 time the observational uncertainties dominate the model
 uncertainties;  the model
 presented here is adequate for an estimate of planet bulk composition.
 In any event, the main results of our work described in the following subsections 
 despite any model uncertainties. 
 
It is possible to rule out parts of the ternary diagram as being
physically unplausible \citep[e.g.,][]{vale2007}. This is based on the initial composition of the 
protoplanetary nebula and on planet differentiation.
For example, a pure iron planet is unlikely to exist, because removing all
of the mantle would be difficult. A pure water planet is also unlikely to exist.
Where water ice forms, so do silicate-rich and iron-rich materials, making planet accretion
of pure water unlikely. We prefer to leave
the omission of parts of the ternary diagram to users of the code, because
in exoplanets surprising exceptions to the "rules" of planet characteristics are not uncommon.

\subsection{Spacing, Shape, Direction, and Rotation of Curves on
the Ternary Diagram.}

We now turn to a discussion of the spacing, direction, and shape of
the curves in the ternary diagrams. A quantitative and qualitative
description of these is a main point of this paper.  We emphasize that
each curve shown in our Figure~\ref{fig:ternmult} diagrams represents
a different mass and radius. The curves to the lower right are more
dense, as they have a higher mass and lower radius than the curves
moving to the upper left.

All behavior results from the equations of state of the materials,
and, in some cases, how they behave differently under pressure.

We begin with an equation that we use repeatedly in this section.  We
consider the simplified case that the planet core has a uniform
density, where $ {\bar{\rho}_{Fe}}$ is the average density of Fe in
the core, ${\bar {\rho}_{MgSiO_3}}$ is the average density of silicate
in the mantle, and ${\bar {\rho}_{H_2O}}$ is the average density of
water in the outer water layer.  We then have
\begin{equation}
\label{eq:basic}
\frac{4}{3} \pi R_p^3 = \left[
 \frac{\alpha M_p}{\bar {\rho}_{Fe}} +
 \frac{\beta M_p}{\bar{\rho}_{MgSiO_3}} +
 \frac{\gamma M_p}{\bar{\rho}_{H_2O}}
 \right],
\end{equation}
where $ \alpha + \beta + \gamma = 1$.

\subsubsection{Spacing}

The iso-mass iso-radius curves for adjacent
curves with equal differences of mass and radius  have uneven spacing on a ternary
diagram. This spacing is generally smaller in the lower right part of the ternary diagram (high iron fraction region) than in the upper left part (high silicate or water fraction region). 
The curves in the lower right  part of the diagram have a higher density (higher mass
and smaller radius) than the curves found on the upper left part of the diagram. The density is not a linear function of both mass and radius hence we do not expect equal 
spacing on the ternary diagram. We can, however, give both a quantitative and qualitative 
explanation of the uneven spacing.

We can provide a quantitative description, beginning with equation~(\ref{eq:basic}),
but using  $\alpha = 1 - \beta - \gamma$ to get
\begin{equation}
\frac{4}{3} \pi R_p^3 = M_p \left[
 \frac{(1 - \beta - \gamma) }{\bar {\rho}_{Fe}} +
 \frac{\beta }{\bar{\rho}_{MgSiO_3}} +
 \frac{\gamma}{\bar{\rho}_{H_2O}}
 \right]
\end{equation}

We use the fact that the curves on the ternary diagram
are almost perpendicular to the water side of the ternary diagram,
and therefore set the silicate mass fraction $\beta = 0$ for our discussion. In other words, the
distance (separation) between the points 
produced by the intersection of the isomass-isoradius curves and the water axis
is a good representation 
of the spacing between the curves throughout the ternary diagram,
\begin{equation}
\label{eq:basic2}
\frac{4}{3} \pi R_p^3 = M_p 
\left[ \frac{1}{\bar {\rho}_{Fe}} +
 \gamma \left( 
 \frac{1}{\bar {\rho}_{H_2O}} -
 \frac{1}{\bar{\rho}_{Fe}} \right)
 \right].
\end{equation}

We define the following constants:
\begin{equation}
A_1 = \frac{1}{\bar {\rho}_{Fe}}
\end{equation}
and
\begin{equation}
A_2 = \frac{1}{\bar {\rho}_{H_2O}} - \frac{1}{\bar {\rho}_{Fe}}.
\end{equation}
Equation~(\ref{eq:basic2}) then becomes
\begin{equation}
\gamma A_2 = \frac{4/3 \pi R_p^3}{M_p} - A_1.
\label{eq:basic3}
\end{equation}
We note that $A_1>0$ and $A_2 > 0$.

Now we proceed to take the derivative of
equation~({\ref{eq:basic3}) with respect to the
change of mass ($dM_p$) and change of radius ($dR_p$).
\begin{equation}
A_2 d \gamma = \frac{4 \pi}{3} \left[
\frac{3 R_p^2}{M_p} dR_p
- \frac{R_p^3}{M_p^2}dM_p
\right].
\end{equation}
We can also rewrite equation~({\ref{eq:basic3}) in terms
of the overall average density of the planet ($\bar{\rho}$),
\begin{equation}
\gamma A_2 = \frac{1}{\bar \rho} - A_1,
\end{equation}
and the corresponding derivative relative to the overall average density ($\bar{\rho}$)
\begin{equation}
A_2 d\gamma  = -\frac{1}{\bar \rho^2} d {\bar \rho}.
\label{eq:basic4}
\end{equation}

This leads to our quantitative understanding, where we first recall
that $d \gamma$ is the water fraction spacing on the ternary diagram.
The lower right part (Fe-rich) of the ternary diagram is where the
average density ($\bar \rho$) of a planet is high. This implies (for
the same $d \bar \rho$) $d \gamma$ is small (since $\bar \rho$ is in
the denominator).  In the upper left region of the ternary diagram
(water-rich) the average planet density is smaller than a planet
located in the lower right part of the diagram, and therefore $d
\gamma$ is larger.

Qualitatively, to have wider spacing an increasing water fraction is
needed.  In other words, towards the upper right part of the ternary
diagram, for the same density difference more water than iron must be
replaced.

We note that the spacing is predominantly the 
result of the nonlinearity of equations~(\ref{eq:basic3}) and (\ref{eq:basic4})
and the average density of each compositional layer, 
not of  any $P$-$\rho$ properties of the EOS (i.e.,
how materials condense under high pressure). This statement is correct
under the assumption of a single material  for each layer in the planet 
(in our case for iron, silicate, and water). The assumption 
that the average density within each layer does not change significantly from curve to curve (for example, from the 1-$\sigma$ curve to the 3-$\sigma$ curve for the case $M_p=2 M_\oplus$ and $R_p=1.5 R_\oplus$)  is reasonable for a ternary diagram that spans only a small 
mass and radius range.

\subsubsection{Shape, Direction, and Rotation}

To explain the shape and direction of the curves in the ternary diagrams
we start by explaining the slope of the curves in the Cartesian diagram.
In other words, we are aiming for an expression of $d \alpha / d \beta$.

We start with  a different form of equation~(\ref{eq:basic}),
\begin{eqnarray}
\alpha \left[
\frac{1}{\bar \rho_{H_2O}} - \frac{1}{\bar \rho_{Fe}} \right]
+
\beta \left[
\frac{1}{\bar \rho_{H_2O}} - \frac{1}{\bar \rho_{MgSiO_3}} \right]
= \nonumber \\
\frac{1}{\bar \rho_{H_2O}} - \frac{4 \pi R_p^3}{3 M_p}.
\end{eqnarray}
We now differentiate this equation with respect to $\alpha$ and $\beta$ to find
\begin{equation}
d\alpha \left[
\frac{1}{\bar \rho_{H_2O}} - \frac{1}{\bar \rho_{Fe}} \right]
+
d\beta \left[
\frac{1}{\bar \rho_{H_2O}} - \frac{1}{\bar \rho_{MgSiO_3}} \right]
\approx 0,
\end{equation}
based on the assumption that the average density is changing slowly
with respect to the change in composition.

We write the slope of the curves in the Cartesian diagram
\begin{equation}
\label{eq:slopecartesian}
\left | \frac{d \beta}{d \alpha} \right | = 
\frac{ \left[ 1 - \frac{\bar \rho_{H_2O}}{\bar \rho_{Fe}} \right]}
{ \left [1 - \frac{\bar \rho_{H_2O}}{\overline \rho_{MgSiO_3}} \right]}
\end{equation}
We can see that 
\begin{equation}
\left | \frac{d \beta}{d \alpha} \right | > 1,
\end{equation}
assuming that ${\bar \rho_{H_2O}}  < {\bar \rho_{MgSiO_3}} < {\bar \rho_{Fe}}$.


It can be shown that the slope in the the ternary diagram 
has a positive connection to the slope in the Cartesian diagram.
Using the the equations that convert the Cartesian coordinates
to coordinates on the ternary diagram (equations~(\ref{eq:coordconversiona}) and
(\ref{eq:coordconversionb})) we find that
\begin{equation}
\label{eq:slopeternary}
\left | \frac{d \beta_{ternary}}{d \alpha_{ternary}} \right | = 
\sqrt{3}
\frac{\left[ \left | \frac{d \beta}{d \alpha} \right | - 1 \right]}
{\left[ \left | \frac{d \beta}{d \alpha} \right | + 1\right]}.
\end{equation}

We can now go on to describe the direction, shape, and rotation in the
Cartesian diagram based on equation~(\ref{eq:slopecartesian}), with
the knowledge that the same qualitative behavior will appear in the
ternary diagrams.  We first emphasize that the slope of an isomass,
isoradius curve on the ternary diagram describes adding and removing
mass of the different species.

We begin with a qualitative explanation of the direction of the curves
on the ternary diagram---why each curve goes from the lower left to
the upper right.  This is largely a coincidence in the different
values of ${\bar \rho_{H_2O}}$, ${\bar \rho_{MgSiO_3}}$, and ${\bar
\rho_{Fe}}$.  The
coincidence lies in the fact that iron is more dense and water less dense
than silicate---and for zero-pressure densities, an equal
mass of iron and water combined densities are roughly 
similar to the silicate density. At the lower left of the ternary diagram in
Figure~\ref{fig:ternproj}, the silicate mass fraction is 80 percent,
the iron mass fraction is zero and the water mass fraction is 20
percent. As the silicate fraction decreases, a combination of equal parts iron and
water must be added to maintain the same overall planet mass and
radius.  This description is consistent with the direction
of the curves in the ternary diagrams. 

More quantitatively, from equation~(\ref{eq:slopecartesian}),
we can take the zero-pressure densities of Fe, H$_2$O, and MgSiO$_3$
to find the slope of the curve on the cartesian diagram
\begin{equation}
\label{eq:slopecart2}
\left | \frac{d \beta}{d \alpha} \right | = 2,
\end{equation} 
and from equations~(\ref{eq:coordconversiona}) and
(\ref{eq:coordconversionb}),
\begin{equation}
\left | \frac{d \beta_{ternary}}{d \alpha_{ternary}} \right | = \frac{1}{\sqrt{3}}.
\end{equation} 
We can consider removing a fixed amount
of iron mass, e.g., 1 gram. According to equation~(\ref{eq:slopecart2}),
2 grams of silicate must be added. 
For mass balance (because each curve on the ternary
diagram represents a planet of fixed mass and fixed radius), 1 gram of water must be removed. 
The direction of the curves on the ternary diagram do correspond
to removing roughly equal masses of iron and water for
every mass of silicate added.

{\it Rotation} 

In Figure~\ref{fig:ternmult}a--d (where a through d are in
order of increasing mass), we see that for more massive
planets, the isomass-isoradius curves are rotated. In other words, the
slope of the curve on the ternary diagram increases for increasing mass.

We again return to equation~(\ref{eq:slopecartesian}).
With increasing planet mass, each of
 ${\bar \rho_{H_2O}}$,  ${\bar \rho_{MgSiO_3}}$, and $ {\bar \rho_{Fe}}$
 changes. Because iron is in the core, it suffers
 more compression than the silicate mantle or water icy layer. 
 In other words,  $ {\bar \rho_{Fe}}$
 must increase more than ${\bar \rho_{H_2O}}$ and  ${\bar \rho_{MgSiO_3}}$
 as the planet mass increases.
Therefore, the numerator of equation~(\ref{eq:slopecartesian})
gets larger and thus the slope of the curve
increases. The rotation is counter clockwise with increase of mass.

{\it Slope of the Curves}

We now turn to discuss the slope of each isomass-isoradius curve.
We see from Figures~\ref{fig:ternunc} and \ref{fig:ternmult} that the
slope of each isomass-isoradius curve for a fixed mass and radius is greater in the lower
left part of the curve than in the upper
right part of the curve. 

The slope is due to the
differential compression of water, silicate and iron under pressure.
This slope is again explained by
equation~(\ref{eq:slopecartesian}).  At the lower left the slope is
smaller; this is the silicate-rich region of the ternary
diagram. There is more silicate and less iron and water.  The ${\bar
\rho_{MgSiO_3}}$ will increase as it gets compressed in the inner part
of the planet.  This causes the slope in
equation~(\ref{eq:slopecartesian}) to get smaller.  In contrast, at
the upper right, there is little silicate, but more water and more
iron.  The silicate and water are less compressed, but the iron is
more compressed, making the slope increase.

For a conceptual explanation, first recall the idea described above that
removing 2 grams of silicate can be compensated by adding approximately
1 gram of water and 1 gram of iron.
In a silicate-rich planet (lower left of the ternary diagram),
silicate is compressed. For a massive planet, this compression makes
the silicate density closer to iron's density than in the uncompressed case.
Therefore, removing a fixed mass of silicate requires much more iron than water to be added,  for a fixed
mass and radius.
 
In contrast, along the upper right part of the ternary diagram, the
planet is iron-rich or water-rich. The slope of an isomass-isoradius
curve is steeper than a curve in the lower left part of the ternary
diagram. For an isomass-isoradius curve in the upper part of the
ternary diagram, the planet has more iron, the iron is very compressed
(There is less silicate and the silicate is overall less compressed
compared to a planet in the lower left of the ternary diagram). If a
fixed mass of silicate is removed, more water than iron must be added
to compensate for the density of compressed iron. The density of water
does not change much, because water is always the outer layer and thus
the least compressed.


As a qualitative explanation, if the average density of each of the
three layers remains constant, then an isomass-isoradius curve should
always be a straight line in either the Cartesian or ternary
representation. The curvature in a given isomass-isoradius curve
appears because of the compression of material under pressure, which
is a property of the EOS.

\subsection{Shape and Direction of an Isoradius Surface in our 3D Representation}

We return to the 3D representation of the relationship between mass,
radius and composition shown in Figure~\ref{fig:isorisom}.  An
isoradius surface (red oblique surface) is shown for $R_p = 1.7
R_\oplus$. An isomass surface (one of the blue flat planes) is shown for
$M_p=4 M_\oplus$. At the bottom tip of the isoradius surface, the
silicate mass fraction = iron mass fraction = 0, and the planet is
composed of 100\% water.  This is the minimum mass for this
radius. The isoradius surface also has a maximum mass, reached by a
composition of 100 percent iron.

For the same radius, if either the silicate or iron mass fraction is
increased, the planet mass must also increase. This is because iron
and silicate are denser than water ice. We further note that for
an increase in iron mass fraction, the mass of the planet must
increase more steeply than for an increase in the silicate mass
fraction. This is seen by the different length and shape "edges" of
the isomass surface in the iron-mass plane and silicate-mass plane
in Figure~\ref{fig:isorisom}.

We now show that the shape of the isoradius surface is concave. 
The isoradius surface can be considered as the isovolume surface, where the volume is the sum of
the core, the silicate mantle, and the water crust. To calculate the total mass of a point on the isoradius curve, we use equation~(\ref{eq:basic}), and since $R_p$ is constant on this surface we can rewrite this equation (with $C$ a constant) as,
\begin{equation}
M_p =  C
\left[
\frac{\alpha}{\bar{\rho_{Fe}}} +
 \frac{\beta}{\bar{\rho_{MgSiO_3}}} +
 \frac{(1 - \alpha - \beta) }{\bar{\rho_{H_2O}}}
 \right]^{-1}.
\end{equation}
The term in brackets is a linear function and its inverse is 
hyperbolic.
For example, if we let $\alpha = 0$ (no iron), then
we have
\begin{equation}
M_p = C \left[
 \frac{1}{\bar{\rho_{H_2O}}}
 - \beta \left( \frac{1}{\bar{\rho_{H_2O}}} -\frac{1} {\bar{\rho_{MgSiO_3}}} \right)
\right]^{-1},
\end{equation}
where
\begin{equation}
C = \frac{4}{3} \pi R_p^3.
\end{equation}
Both terms in the square brackets are positive and 
$ 0 \leq  \beta \leq 1$.
This is in the form
\begin{equation}
M_p = \frac{C}{ C_1 - C_2 \beta},
\end{equation}
where
\begin{equation}
C_1 = \frac{1}{{\bar \rho_{H_2O}}}
\end{equation}
and
\begin{equation}
C_2 = \frac{1}{{\bar \rho_{H_2O}}} - \frac{1}{{\bar \rho_{MgSiO_3}}}
\end{equation}
We have  $ C_1 > C_2 > 0$.
This is the form of a concave hyperbola,
because, taking the first derivative we find
\begin{equation}
\frac{dM_p}{d\beta} = \frac{C C_2}{\left ( C_1 - C_2 \beta \right)^2}
\end{equation}
The slope increases as $\beta$ increases  because
$C_1 - C_2 \beta$ decreases but is always greater than zero.

Although we have made the simplification
that the average densities of each layer remain constant,
 our qualitative description holds because
$  {\bar \rho}_{H_2O} < {\bar \rho}_{MgSiO_3} < {\bar \rho}_{Fe}$ 
as long as the planet is differentiated into layers of increasing
density towards the planet center.

A similar argument shows that the total iron fraction
vs. total mass is also a hyperbola, making the whole
isoradius surface concave.

\section{Summary and Conclusion}

An ambiguity in an exoplanet interior composition remains for
any planet with a measured mass and radius, no matter how 
precisely measured. We can accept this ambiguity and quantify it with the aid
of ternary diagrams \citep{vale2007}. We have presented ternary diagrams
for a single planet of fixed mass and radius, for a planet composed of 
an iron core, a silicate mantle, and a water ice outer layer. Our ternary
diagram presentation includes observational  uncertainties. 
We have provided a publically available computer code to generate a ternary
diagram for a given input mass, radius, and observational uncertainties.

In addition to presenting ternary diagrams for fixed mass and radius, 
we showed their origin from a 4D database ($M_p$, $R_p$, iron mass-fraction $\alpha$,
and the silicate mass fraction $\beta$; recall that the water-ice mass fraction
 $\gamma = 1 - \alpha - \beta$). We further described the shape and direction of
 the composition curves on a ternary diagram .
 
We conclude with the sentiment that in order to fully
understand the interior structure of an exoplanet, 
a third measurement beyond planet mass and radius is required.

\acknowledgements{We thank the MIT John Reed Fund for supporting an
undergraduate research opportunity for L. Z.}

\bibliography{planets}

\begin{figure}
\plotone{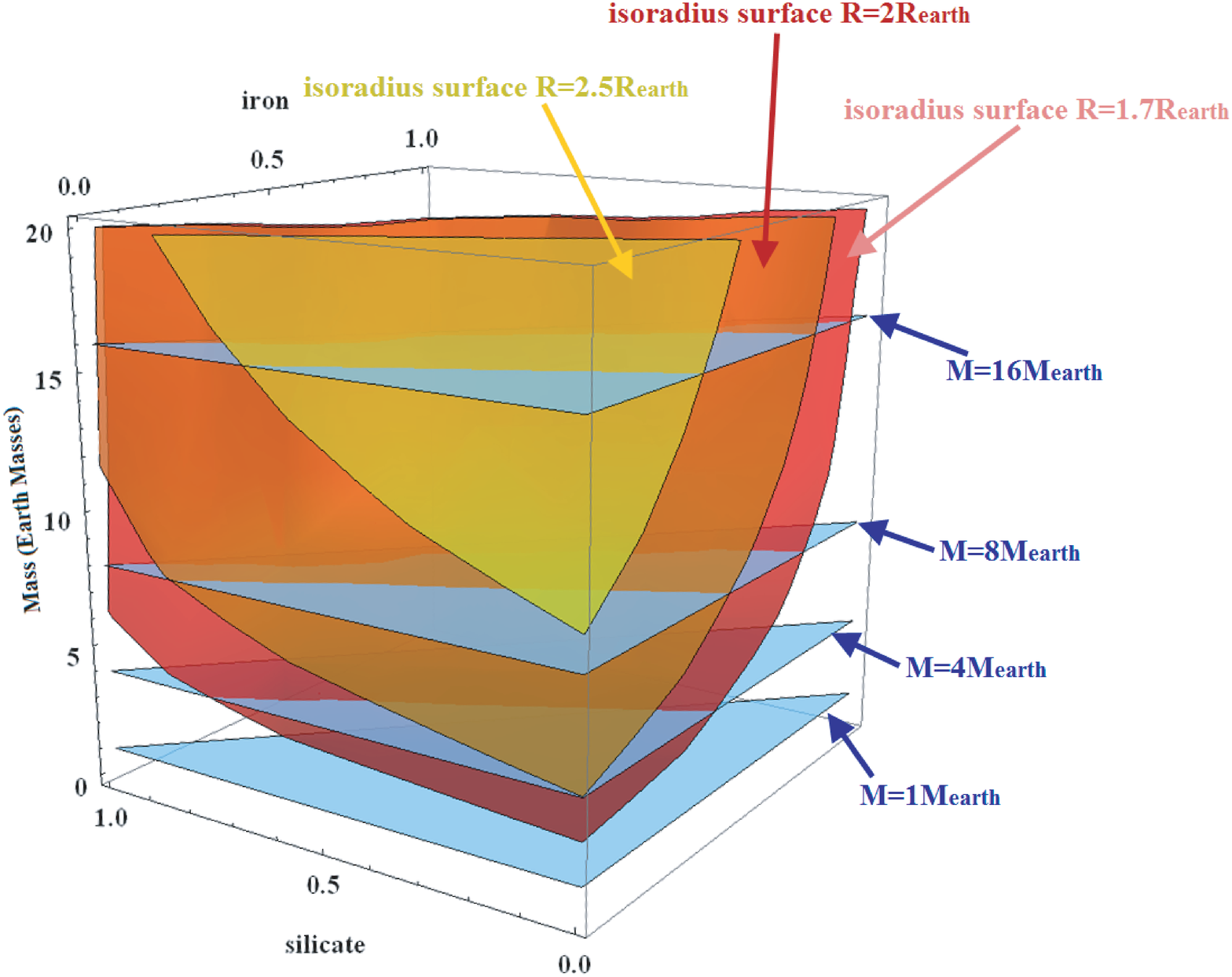}
\caption{The 3D representation of planet mass, planet radius, and
composition expressed as mass fractions of iron ($\alpha$) and
silicate ($\beta$).  The $x$-axis is the iron mass fraction and the
$y$-axis is the silicate mass fraction. The water mass fraction comes
from $\alpha + \beta + \gamma = 1$. The $z$-axis is the total planet
mass in Earth masses.  The planet radius is not indicated, except by
three separate surfaces of constant radius.  An isomass and isoradius
surface intersect with a curve in a 2D iron mass-fraction vs. silicate
mass fraction Cartesian diagram (Figure~\ref{fig:cartproj}).}
\label{fig:3D}
\label{fig:isorisom}
\end{figure}

\begin{figure}
\plotone{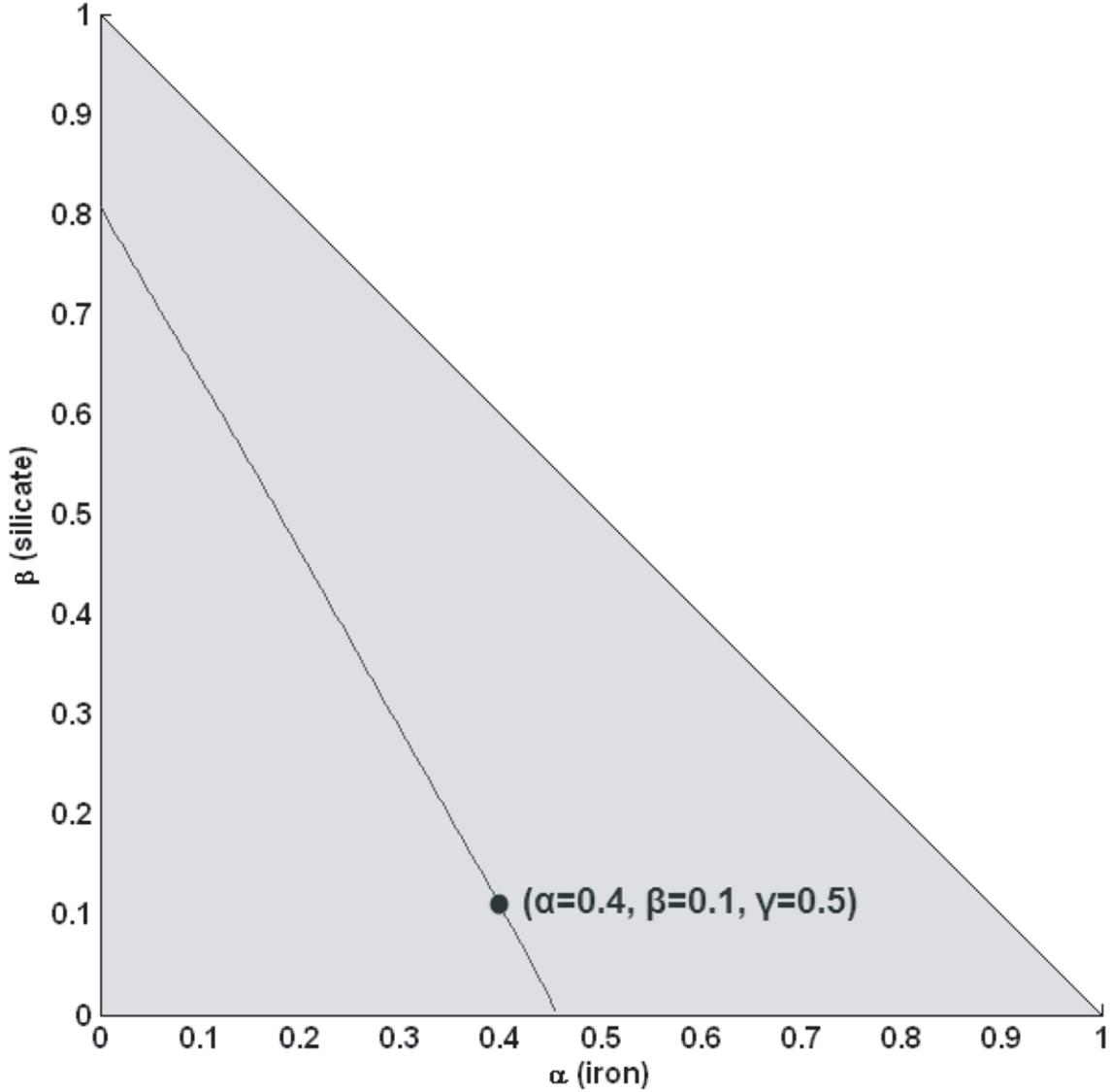}
\caption{The Cartesian diagram for silicate mass fraction ($\beta$)
vs. iron mass fraction ($\alpha$) for a planet with $M_p = 4
M_{\oplus}$, $R_p = 1.7 R_{\oplus}$ (Note that the water mass fraction
$\gamma = 0.4$ since $\gamma = 1 - \alpha - \beta.$) A planet of a
fixed internal composition is represented as a point in the grey
shaded region (only compositions in the grey shaded region are
allowed.)  With a given planet mass and radius there is not a unique
interior composition; the allowed compositions are described by the
curve.  The curve originates from the intersection of the isomass
surface and the isoradius surface in the 3D representation
(Figure~\ref{fig:isorisom}).}
\label{fig:cart1}
\label{fig:cartproj}
\end{figure}

\begin{figure}
\plotone{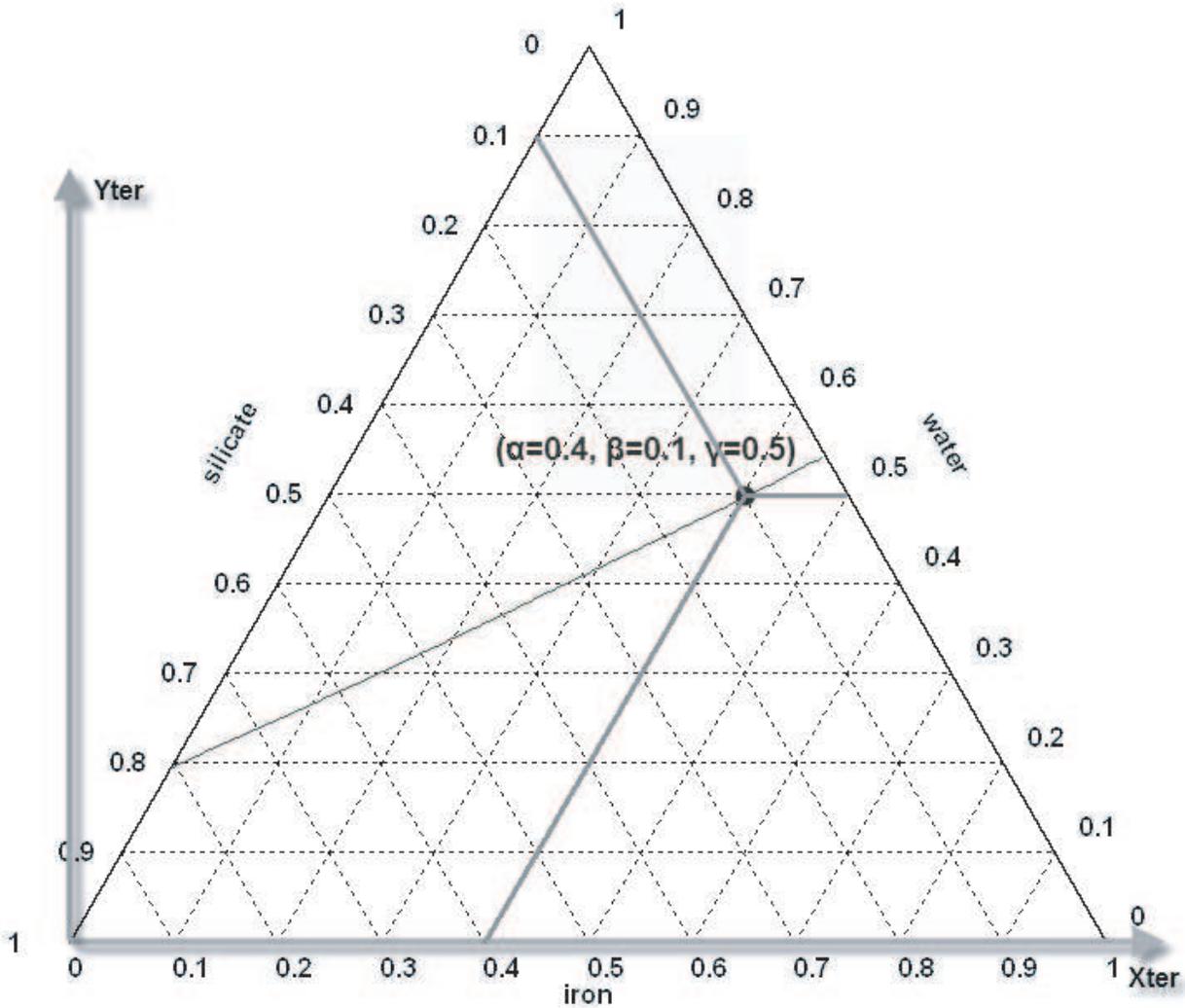}
\caption{A ternary diagram for a planet of the same mass and radius as
shown in the corresponding Cartesian diagram in
Figure~\ref{fig:cart1}.  The light grey $x$ and $y$ axes are shown to
illustrate how this ternary diagram relates to a Cartesian diagram;
$x_{ternary}$ and $y_{ternary}$ are the coordinates transformed from
the Cartesian coordinates.  (See \S~\ref{sec-convert} for the conversion
equations.)  Any point on the curve is a possible combination of iron,
silicate, and water that will result in a planet with with $M_p = 4
M_{\oplus}$ and $R_p = 1.7 R_{\oplus}$. This Figure also illustrates
how to read a ternary diagram. Consider a triangle oriented such that
the value 1 (of a given material) is at the triangle's apex. The mass
fraction of that given material can be read off of a horizontal line
that is perpendicular to the line connecting the triangle's apex with
the triangle base.}
\label{fig:terndef}
\label{fig:ternproj}
\end{figure}

\begin{figure}
\plotone{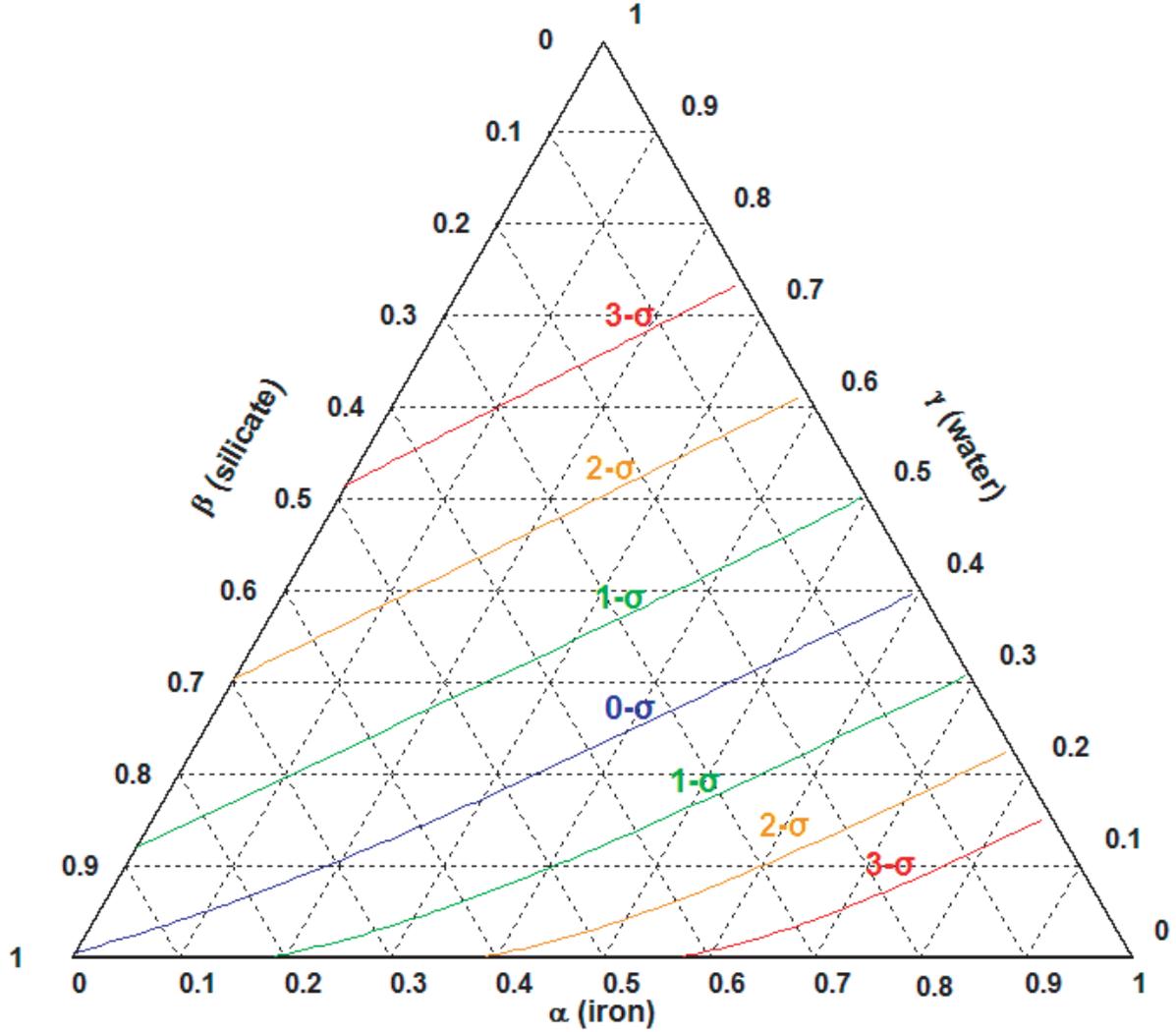}
\caption{ A ternary diagram including the mass and radius
uncertainties for a planet of a fixed mass and radius. This example is
for a planet with $M_p = 10 \pm 0.5 M_{\oplus} $ and $R_p = 2 \pm 0.1
M_{\oplus}$, showing the 1-, 2- and 3-$\sigma$ uncertainty curves.
This Figure shows the continuous distribution of possible combinations
of iron, silicate, and water with increasing uncertainties according
to the color bar. Notice that considering the 3-$\sigma$ uncertainties
almost the entire ternary diagram is covered---in other words there is
no constraint on the planet internal composition. See text for a
discussion of the direction and spacing of the curves.  }
\label{fig:contour}
\label{fig:ternunc}
\end{figure}

\begin{figure}
\plotone{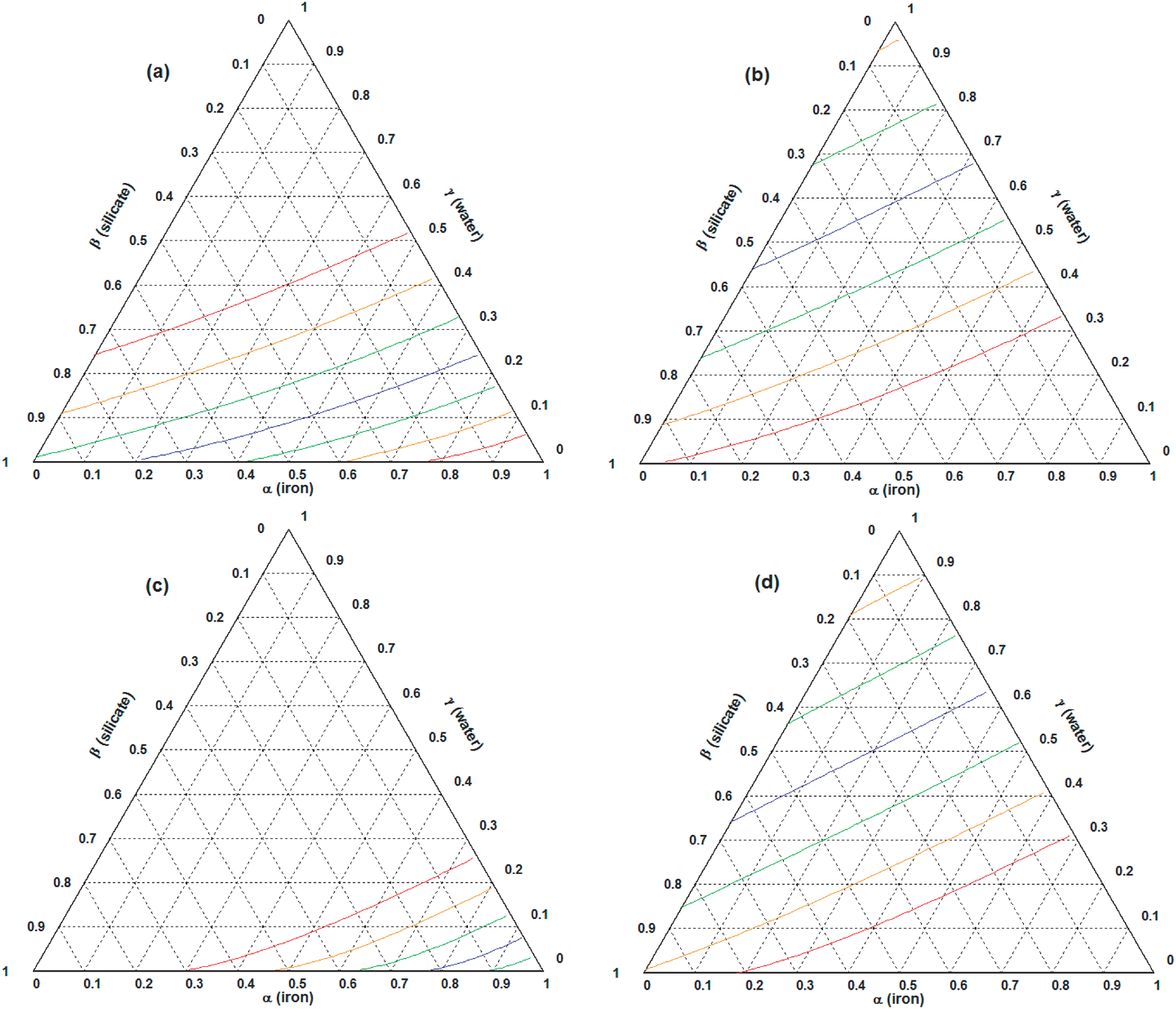}
\caption{Ternary diagrams including 5\% mass and radius uncertainties
for planets of fixed mass and radius. Panel a) $M_p = 1 \pm 0.05 M_{\oplus}$ 
and $R_p = 1 \pm 0.05 M_{\oplus}$. Panel b) $M_p = 2 \pm 0.1 M_{\oplus}$ 
and $R_p = 1.5 \pm 0.075 M_{\oplus}$. Panel c) $M_p = 8 \pm 0.4 M_{\oplus}$ 
and $R_p = 1.5 \pm 0.075 M_{\oplus}$. Panel d) $M_p = 16 \pm 0.8 M_{\oplus}$ 
and $R_p = 2.5 \pm 0.125 M_{\oplus}$. See text for a discussion of the
direction and spacing of the curves.  }
\label{fig:ternmult}
\end{figure}

\end{document}